\documentclass[
    ,final            
  ]
  {aipproc}

\layoutstyle{6x9}

\usepackage{pgf}

\newcommand\dd{\mathrm{d}}
\newcommand\tr{\mathrm{tr}}
\newcommand\M{\mathcal{M}}
\newcommand\e{\mathrm{e}}
\newcommand\ther{\mathrm{th}}
\newcommand\be{\beta}
\newcommand\Un{\mathrm{I}}
\newcommand\G{\mathcal{G}}
\begin{document}

\title{Thermodynamic and quantum entropy gain of frame averaging}

\classification{03.67.-a, 03.65.Ta}
              
\keywords      {irreversibility, relative entropy, frameness, twirl}

\author{Lajos Di\'osi}{
  address={Research Institute for Particle and Nuclear Physics\\
           H-1525 Budapest 114, P.O.Box 49, Hungary}
}

\begin{abstract}
We are discussing a universal non-unitary map M subsequent to a generic unitary map U, 
whose von Neumann entropy gain coincides with the calculated thermodynamic entropy production. 
For many-body quantum reservoirs we prove that M can be the averaging over all translations 
of the spatial frame. Assuming the coincidence of microscopic and macroscopic entropy productions 
leads to a novel equation between entropy gain of frame averaging and relative entropy.
Our map M turns out to coincide with the older one called twirl, used recently in the theory
of quantum reference frames. Related results to ours have been obtained and we discuss some of
them briefly. Possible relevance of frame averaging (twirling) for real world irreversibility 
is mentioned.      
\end{abstract}

\maketitle
In a sequence of works \cite{Dio02,DioFelKos06,Dio07}, 
we studied how we would generate microscopic entropy in simple models
where the amount of thermodynamic entropy production is also calculable. By requiring the
identity of both the microscopic and thermodynamic entropy productions, we were led to the
concept of `graceful' irreversible map and to a mathematical conjecture:
the von Neumann entropy generated by the map is equal to the relative von Neumann entropy
of the state before and after the map. For a special case, the conjecture was proven by 
Csisz\'ar, Hiai and Petz, also it got an application to a certain quantum channel capacity 
problem \cite{CsiHiaPet07}. 
In the same years, and independently of our research, several works on
the notion and quality of reference frames appeared, cf. e.g. \cite{Vacetal08} by Vaccaro et al.,
\cite{Gouetal09} by Gour et al. They defined the measure of frameness, i.e.: of
the goodness of a reference frame, as the entropy generated by `twirl'. This map was used earlier by
Bennett et al. \cite{Benetal96}, and our `graceful' irreversible map happens to coincide with it.

I am going to recapitulate the unpublished work \cite{Dio07} which
generalizes \cite{Dio02} and \cite{DioFelKos06} for bulk quantized systems, 
I'll furthermore include a quick reference to the mentioned concept of frameness. 

\section{Entropies: Von Neumann $S$ vs thermodynamic $S^\ther$}
Consider a certain homogeneous equilibrium reservoir at temperature $T=1/\be$ 
and volume $V$, with density matrix
\begin{equation}\label{rhob}
\rho_\be=\frac{1}{Z_\be}\e^{-\be H}~,
\end{equation}
where $H$ is the Hamiltonian. In the thermodynamic limit, 
the reservoir's von Neumann (microscopic) entropy 
$S(\rho_\be)=-\tr(\rho_\be\log\rho_\be)$ coincides with the
thermodynamic (macroscopic) entropy $Vs(\be)$:  
\begin{equation}
\lim_{V\rightarrow\infty}\frac{S(\rho_\be)}{V}=s(\be)~.
\end{equation}
This relationship assures that the microscopic theory can recover the macroscopic
thermodynamics for equilibrium states. For non-equilibrium states, however, 
the general proof is missing. The present work enforces the coincidence of
von Neumann and thermodynamic entropy productions, respectively, by constructing
a certain `graceful' irreversible map $\rho\rightarrow\M\rho$ which is  
mathematically simple, and may reflect some universal features of the true mechanism
of irreversibility.   

\section{$S$ and $S^\ther$ in non-equilibrium}
Assume we switch on a certain external field to act during a limited period 
and in a limited spatial region. 
When the field is switched off, 
the state of the reservoir becomes the unitary transform $\rho_\be'=U\rho_\be U^\dagger$
of $\rho_\be$, preserving the entropy: 
\begin{equation}\label{SrhobSrhobpr}
S(\rho_\be')=S(\rho_\be)~.
\end{equation}
The field performs a certain work 
$W=\tr(H\rho_\be')-\tr(H\rho_\be)$ on the reservoir. In our model, we \emph{suppose} 
that the entire work $W$ becomes dissipated in the reservoir hence the thermodynamic entropy
production is just $\Delta S^\ther=\be W$ according to standard thermodynamics.
Let us invoke the definition of the von Neumann relative entropy of two density matrices:
\begin{equation}\label{Klein}
S(\sigma\vert\rho)=\tr[\sigma(\log\sigma-\log\rho)]\geq0~.
\end{equation}
It is a remarkable fact that the macroscopic entropy production $\Delta S^\ther=\be W$
can be expressed as the
microscopic (von Neumann) relative entropy of the initial and final quantum states:
\begin{equation}\label{DS}
\Delta S^\ther=S(\rho_\be'\vert\rho_\be)~.                
\end{equation} 
The derivation is elementary \cite{DioFelKos06,Per93}. We just read $\be H=\log Z_\be+\log\rho_\be$
from eq.~(\ref{rhob}) and substitute it into:
\begin{equation}
\Delta S^\ther=\be W=\tr[\be H(\rho_\be'-\rho_\be)]~,
\end{equation} 
which, due to (\ref{SrhobSrhobpr}), yields $S(\rho_\be'\vert\rho_\be)$. 
One would expect that $\Delta S^\ther$ coincides with the increase of von Neumann entropy in the reservoir 
in the thermodynamic limit, but it does not:
\begin{equation}
\lim_{V\rightarrow\infty}\left[S(\rho_\be')-S(\rho_\be)\right]
\not=\lim_{V\rightarrow\infty}\Delta S^\ther~.
\end{equation}
The l.h.s. is always zero, cf. eq.~(\ref{SrhobSrhobpr}), since $\rho_\be$ and $\rho_\be'$ are unitarily equivalent. 
On the macroscopic end, however, we \emph{assumed}
that the state $\rho_\be'$ differs from $\rho_\be$ irreversibly
because of the dissipated work. This is just the 150-year-old
conflict between microscopic reversibility and macroscopic irreversibility. 
We do not intend to propose any ultimate resolution of this conflict.
We only propose a formal resolution which may not be the real mechanism of
irreversibility, yet it may have something to do with it. The novelty of our
approach is that, at very general model conditions, it incorporates the exact 
value of macroscopic entropy production into the microscopic dynamics of the reservoir. 

\section{A graceful irreverzible map $\M$}
We are searching for a formal irreversible 
(i.e.: non-unitary) map $\M$ which has two attractive features. First,
it is graceful in the sense that it conserves the free dynamics of
the reservoir, i.e.:
\begin{equation}\label{grace}
\M [H , \rho] = [H , \M\rho] 
\end{equation}  
for all $\rho$.
Second, it makes the von Neumann entropy production equal to the thermodynamic
one $\Delta S^\ther$ which the eq.~(\ref{DS}) has re-expressed as the relative entropy:
\begin{equation}\label{thconstr}
\lim_{V\rightarrow\infty}\left[S(\M\rho_\be')-S(\rho_\be)\right]=
\lim_{V\rightarrow\infty}S(\rho_\be'\vert\rho_\be)~.
\end{equation}
Such irreversible map was first considered for the Maxwell-gas in the context
of the Boltzmann's rather than the von Neumann's entropy \cite{Dio02}. 
In ref.~\cite{DioFelKos06} we constructed $\M$ for the case of a chain
of non-interacting Pauli-spins, the rigorous proof of which appeared in 
\cite{CsiHiaPet07}.
In the present work, we are going to suggest $\M$ for the realistic
reservoir like, e.g., a quantum gas or liquid of arbitrary strong 
self-interaction and of arbitrary strong perturbation $U$.

Let $H$ be a second quantized Hamiltonian of our reservoir confined in 
a rectangular box of volume $V$ with periodic boundary conditions.
Introduce the unitary operator $U(x)$ of spatial translation\footnote{
Despite similar notations, don't confuse translations $U(x)$ with the local perturbation $U$ in
$\rho_\be'=U\rho_\be U^\dagger$.}
by the vector $x$. Since $U(x)HU(-x)=H$ for all $x$, also the equilibrium state
(\ref{rhob}) will be translation invariant: $U(x)\rho_\be U(-x)=\rho_\be$.
The locally perturbed state $\rho_\be'$ can not be translation invariant. 
For it, consider the following irreversible (non-unitary) map:
\begin{equation}\label{MrhoxV}
\M\rho_\be'=\frac{1}{V}\int U(x)\rho_\be' U(-x)\dd x~,
\end{equation}
where the integration extends for the volume $V$ of the box.
This map satisfies the eq.~(\ref{grace}) of `gracefulness'. 
We conjecture that also the thermodynamic constraint (\ref{thconstr}) is satisfied.

\subsection{Proof}
To outline the proof, we generalize the rigorous method of ref.~\cite{CsiHiaPet07} heuristically. 
Observe that on the l.h.s. of eq.~(\ref{thconstr}) both $S(\M\rho_\be')$ and $S(\rho_\be')$ diverge
for $V\rightarrow\infty$, 
only their difference will converge. Fortunately, one can use the following identity:
\begin{equation}
S(\M\rho_\be'\vert\rho_\be)=-S(\M\rho_\be')+S(\rho_\be')+S(\rho_\be'\vert\rho_\be)~,
\end{equation}
which is easy to inspect from eq.~(\ref{MrhoxV}) and from the translation invariance of
$\rho_\be$. Hence  eq.~(\ref{thconstr}) is equivalent with
\begin{equation}
\lim_{V\rightarrow\infty}S(\M\rho_\be'\vert\rho_\be)=0~.
\end{equation}
According to the Hiai-Petz lemma \cite{HiaPet91}: 
\begin{equation}\label{HP}
S(\sigma\vert\rho)\leq S_{BS}(\sigma\vert\rho)~,
\end{equation} 
where  $S_{BS}(\sigma\vert\rho)=\tr[\sigma\log(\sigma^{1/2}\rho^{-1}\sigma^{1/2})]$
is the Belavkin-Staszewski relative entropy \cite{BelSta82} which one re-writes 
in terms of the function $\eta(s)=-s\log s$:
\begin{equation}\label{BS}
S_{BS}(\sigma\vert\rho)=-\tr[\rho\eta(\rho^{-1/2}\sigma\rho^{-1/2})]\geq0~.
\end{equation}
Let us chain the inequalities (\ref{Klein}) and (\ref{HP}) for $\sigma=\M\rho_\be'$ and $\rho=\rho_\be$:
\begin{equation}\label{ineq}
0\leq S(\M\rho_\be'\vert\rho_\be)\leq S_{BS}(\M\rho_\be'\vert\rho_\be)=-\tr[\rho\eta(\M E_\be)]~,
\end{equation}
where $E_\be=\rho_\be^{-1/2}\rho_\be'\rho_\be^{-1/2}$ and
\begin{equation}
\M E_\be=\frac{1}{V}\int U(x) E_\be U(-x) \dd x~.
\end{equation}
Like in ref.~\cite{CsiHiaPet07}, one must prove that $\M E_\be=\Un$ which means $\eta(\M E_\be)=0$. 
Then the inequalities (\ref{ineq}) yield $S(\M\rho_\be'\vert\rho_\be)=0$ which completes the proof. 

Rather than embarking on a lengthy
rigorous derivation of $\M E_\be=\Un$, we use heuristic arguments. 
We consider second quantized formalism where
all quantized fields satisfy $A(x,t)=\exp(itH)A(x)\exp(-itH)$. Assume, for concreteness, 
pair-potential that vanishes at distances much bigger than $\ell$. It is plausible to assume that   
perturbations have a maximum speed of propagation. Hence,
at any given time $t$ after the unitary perturbation $\rho_\be'=U\rho_\be U^\dagger$ e.g. around the origin, 
there exists a finite volume of radius $r$ such that 
\begin{equation}
[U,A(x,t)]=0 
\end{equation}  
for all $\vert x\vert>r$ and for all local quantum fields. 
We can write $E_\be$ in the form
\begin{equation}
E_\be=\exp(\be H/2)\exp(-\be UHU^\dagger)\exp(\be H/2)~.
\end{equation}
The Hamiltonian contributions of the field $A(x)$ will be cancelled for all $\vert x\vert\gg\ell$.  
Therefore $[E_\be,A(x)]=0$ for all $\vert x\vert\gg\ell$. Due to the finite speed of propagation,
also the more general relationship 
\begin{equation}
[E_\be,A(x,t)]=0 
\end{equation}  
holds at any given later time $t$ provided $\vert x\vert>r$ where $r$ is finite and grows with $t$ 
at the speed of propagation. Let us take the infinite volume limit $V\rightarrow\infty$!
Since the sub-volume of radius $r$, where $A(x,t)$ does \emph{not} commute with $E_\be$, is finite and since
$E_\be$ is a bounded operator, the averaged operator $\M E_\be$ will commute with all fields $A(x,t)$
for \emph{all} coordinates $x$! Hence $\M E_\be=\lambda\Un$ and 
the identity $\tr(\rho_\be\M E_\be)=\tr(\rho_\be E_\be)=1$ yields $\lambda=1$. 

\section{Realistic versions of $\M$}
One might wish to construct the `graceful' irreversible map $\M$ at less artificial conditions of 
regularization than the finite volume and the periodic boundary (\ref{MrhoxV}).
An equivalent construction can be done on the Hilbert space of an infinite volume reservoir:
\begin{equation}\label{MrhoxR}
\M\rho_\be'=\lim_{R\rightarrow\infty}\frac{1}{8\pi R^3}\int \e^{-\vert x\vert/R}U(x)\rho_\be' U(-x) \dd x~.
\end{equation}
The map $\M$ makes the reservoir \emph{forget} some information that amounts exactly to the macroscopic entropy
production. From the lesson
of our previous works \cite{Dio02,DioFelKos06} we have guessed that the real quantum reservoir would
gracefully forget the \emph{location} of perturbation. (It does not need to forget it immediately; it may
do it at any later time.) Now, let us call $R$ the scale of spatial frame coarse-graining.
In concrete cases, the information loss can be well saturated at some finite scale $R\gg r$.
This feature can become important if we generalize the single-shot concept of $\M$ for the time-continuous 
generation of irreversibility.

There is a further alternative which we just mention. Instead of the spatial frame, the temporal
one can be made forgotten: 
\begin{equation}\label{MrhotT}
\M\rho_\be'=\lim_{T\rightarrow\infty}\frac{1}{T}\int_{-\infty}^0 \e^{t/T}U(-t)\rho_\be' U(t)\dd t~,
\end{equation}
where $U(t)=\exp(-iHt)$. This state is definitely different from the result of spatial averaging (\ref{MrhoxR}).
Nevertheless, we conjecture that for $T,V\rightarrow\infty$ it gains the same entropy (\ref{thconstr}). 
This has never been discussed, although the irreversible map (\ref{MrhotT}) of local equilibrium Gibbs-states 
has long been known in advanced statistical physics \cite{Zub74}. Note that universal coarse-graining of the
temporal frame has appeared in a number of independent models, see four of them in ref.~\cite{Dio05}.  

\section{Frameness}
Suppose we want to label $n$ nodes along a closed chain and we introduce $n$ Pauli spins as a reference frame.
If the spins are independent and identical, their composite state is 
\begin{equation}
\rho=\sigma\otimes\sigma\otimes\sigma\otimes\sigma\otimes\dots\otimes\sigma~.
\end{equation}
Such state is invariant if we shift the labels along the chain, hence this state is useless for a reference frame.
Now we alter just one of the $n$ spins:
\begin{equation}
\rho^\prime=\sigma\otimes\sigma^\prime\otimes\sigma\otimes\sigma\otimes\dots\otimes\sigma~.
\end{equation}
This makes already a better reference frame, yet its usefulness depends on the `distance'
between $\sigma'$ and $\sigma$. We need a certain measure of the frame's usefulness, which we call \emph{frameness}.
The frameness should be zero if the frame's quantum state is invariant for the group of shifts, and the
frameness should increase with the asymmetry of the quantum state.

Instead of the above discrete frame, we can suppose a continuous one to label spatial coordinates. 
Like before, we consider a many-body system with translation invariant Hamiltonian and with periodic boundary conditions 
in a rectangular volume $V$. The Gibbs equilibrium state $\rho_\be$ (\ref{rhob}) is translation invariant, hence its frameness
must be zero. Its local perturbation $\rho_\be'=U\rho_\be U^\dagger$ is translation non-invariant, its frameness must
be positive.

Now we have to define frameness $F(\rho)$ for an arbitrary state $\rho$. 
We can rely on entropic quantities and we can choose a plausible definition.

\section{Twirl $\G$}
As we said, $F(\rho)=0$ if and only if $U(x)\rho U(-x)\equiv \rho$, i.e., when
$\rho$ is translation invariant. If it is not, then we introduce the average of $\rho$ over all translations:
\begin{equation}\label{GrhoxV}
\G\rho=\frac{1}{V}\int U(x)\rho U(-x)\dd x~.
\end{equation}
The map $\G$ is called `twirl' and it can be defined for locally compact groups. 
It fully coincides with our `graceful' irreversible map (\ref{MrhoxV}).
The state $\G\rho$ is translation invariant, and it is a `closest' translation
invariant map of $\rho$. Of course, $F(\G\rho)$ is zero for all $\rho$. Now, what 
could be the frameness $F(\rho)$ of a generic $\rho$?

Following Vaccaro et al. \cite{Vacetal08}, 
let the frameness of $\rho$ be measured by the twirl's (\ref{GrhoxV}) gain of entropy:   
\begin{equation}
F(\rho)=S(\G\rho)-S(\rho)~.
\end{equation}
Gour, Marvian and Spekkens \cite{Gouetal09} showed that the 'closest' translation
invariant state to $\rho$ is the twirled state $\G\rho$. Morerover, they were
able to prove that the corresponding shortest entropic distance $S(\rho\vert\G\rho)$ is identical
to the above defined frameness:   
\begin{equation}\label{FSrhoGrho}
F(\rho)=S(\rho\vert\G\rho)~.
\end{equation} 
So, the informatic measure of frameness is the relative entropy of the twirled
state w.r.t. the state itself.

\section{Closing remarks}
There is an obvious overlap between our results and those in \cite{Vacetal08,Gouetal09}
although the initial motivations are definitely different. We don't go into the depth
of comparison, rather we compare the central mathematical results. 
By postulating a joint model of both thermodynamic and von Neumann entropy gains,
we came to the non-trivial conjecture:
\begin{equation}\label{ours}
\lim_{V\rightarrow\infty}\left[S(\G\rho)-S(\rho)\right]=
\lim_{V\rightarrow\infty}S(\rho\vert\rho_0)~,
\end{equation}
where $U(x)\rho_0 U(-x)\equiv\rho_0$ and $\rho=U\rho_0 U^\dagger$ while 
\begin{equation}
\G\rho=\frac{1}{V}\int U(x)\rho U(-x)\dd x~.
\end{equation}
To be correct, the conjecture was derived and the heuristic proof was done 
for the special case $\rho_0=\rho_\be$. However, I guess the proof might be 
extended for translation invariant states $\rho_0$. 

A similar non-trivial theorem was exactly proved by Gour et al. \cite{Gouetal09}. 
To get the relationship (\ref{FSrhoGrho}), they proved for 
all groups and for all states $\rho$ that the entropy gain of twirling is 
identical with the relative entropy of the twirled state w.r.t. the state itself:
\begin{equation}\label{theirs}
S(\G\rho)-S(\rho)=S(\rho\vert\G\rho)~.
\end{equation}
Despite its similarity to our result (\ref{ours}), theirs
is different. Apparently, our conjecture and theorem is asymptotic,
it needs the infinite volume limit. Further investigations may perhaps show
intrinsic relationships between (\ref{ours}) and (\ref{theirs}) together with
their underlying physics. 

While the merit of the theorem (\ref{ours}) is independent of the
validity of our model for real physics, our model and results shed 
more light on the physical mechanism of microscopic irreversibility, 
i.e., on Nature's graceful way to \emph{forget} microscopic data.
Nature would gracefully produce irreversibility just by twirling
our reference frames (or, equivalently, by twirling matter). 
Then Nature is producing the observed thermodynamic irreversibility --- 
at least in our calculable models. Whether this is the real and ultimate 
way for Nature to 'forget' microscopic data remains an open question.

\begin{theacknowledgments}
The author is grateful for support of
the Hungarian Scientific Research Fund (Grant T75129),
the SA/Hu Agreement on Cooperation in Science and Technology,
and the organizers of Quantum Africa 2010.
\end{theacknowledgments}

\end{document}